\documentclass[12pt]{article}
\def\appendix{\par
 \setcounter{section}{0}
 \setcounter{subsection}{0}
 \def\@secapp{Appendix}
 \def\thesection{\Alph{section}}}
\usepackage{bm}
\usepackage{amsmath,amssymb}
\usepackage{amsthm}
\usepackage{mathrsfs}
\usepackage{graphicx}
\theoremstyle{plain}
\newtheorem{Th}{Theorem}[section]
\newtheorem{lemma}{Lemma}[section]
\newtheorem{corollary}{Corollary}[section]
\newtheorem{propo}{Proposition}[section]
\newtheorem{definition}{Definition}[section]
\newtheorem{rem}{Remark}[section]
\newtheorem{algorithm}{Algorithm}[section]

\newcommand{\cC}{{\cal C}}
\newcommand{\cE}{{\cal E}}

\newcommand{\cI}{{\cal I}}

\newcommand{\cL}{{\cal L}}
\newcommand{\cM}{{\cal M}}

\newcommand{\cR}{{\cal R}}
\newcommand{\cS}{{\cal S}}
\newcommand{\cT}{{\cal T}}

\newcommand{\vertices}{{V}}
\newcommand{\avertex}{{v}}

\date{July, 2006}

\begin{document}
\title{Boundary cliques, clique trees and perfect sequences of maximal
cliques of a chordal graph}  

\author{
Hisayuki Hara\\
        Department of Geosystem Engineering\\
University of Tokyo\\
and \\
Akimichi Takemura\\
Graduate School of Information Science and Technology\\
 University of Tokyo}
\maketitle

\begin{abstract}
 We characterize clique trees of a chordal graph in 
 their relation to simplicial vertices and perfect sequences of
 maximal cliques.  
 We investigate boundary cliques defined by Shibata\cite{Shibata} 
 and clarify their relation to endpoints of clique trees. 
 Next we define a symmetric binary relation between the set of clique
 trees  and the set of perfect sequences of maximal cliques. We describe
 the 
 relation as a bipartite graph and prove that the bipartite graph is
 always connected.
 Lastly we consider to characterize chordal graphs from the aspect
 of non-uniqueness of clique trees.
\end{abstract}


\section{Introduction}
Chordal graphs  are useful for many practical problems.
For example they arise in the context of sparse linear systems
(Rose\cite{Rose}), relational data bases (Bernstein and
Goodman\cite{BG}), positive definite completions (e.g.\ Grone et
al.\cite{Grone}, Fukuda et al.\cite{Fukuda-etal}, Waki et
al.\cite{Waki-etal}). In statistics, graphical models have
received increasing attention (e.g.\ Whittaker\cite{Whittaker},
Lauritzen\cite{Lauritzen}).  
The decomposable graphical models determined by chordal graphs are
particularly convenient and
have been extensively studied by many authors 
(e.g.\ Dobra\cite{Dobra}, 
Geiger, Meek and Sturmfels\cite{Geiger-Meek-Sturmfels},
Hara and
Takemura\cite{H-T-2}, \cite{H-T-3}). 
In view of these applications it is important to study properties of
chordal graphs.

In this article we focus on characterizations of clique trees for
a chordal graph. 
A clique tree is an intersection graph representation of a chordal graph
and in general there are many clique trees for a chordal graph.
Clique trees are very important from the algorithmic point of view for 
many techniques based on chordal graphs. 
They have been used to solve domination problems on directed
path graphs (Booth and Johnson\cite{Booth}).
They also provide efficient algorithms for probability propagation in 
graphical models (e.g.\ Jensen\cite{Jensen}). 

The purpose of this article is to characterize the set of 
clique trees in three  ways. 
We first address properties of boundary cliques defined by 
Shibata\cite{Shibata} which form an important subclass of simplicial
cliques. 
Shibata\cite{Shibata} showed that if a maximal clique $C$ is an endpoint 
of some clique tree, then $C$ is a boundary clique. 
We show that the converse of this fact holds and give some
characterizations of endpoints of clique trees by using the notion of
boundary cliques. 
In this paper we also use an alternative terminology and 
call a boundary clique {\em  simply separated}, because
a boundary clique meets a single  minimal vertex separator.
The characterization of endpoints of clique trees is essential
for proving theoretical facts on chordal graphs by induction on the
number of maximal cliques. 

Secondly we consider the relation between the set of clique trees and
the set of perfect sequences of maximal cliques.
Lauritzen\cite{Lauritzen} presents two (randomized) algorithms, one of which
generates a clique tree from a perfect sequence given as an input and
the other generates a perfect sequence of maximal cliques from a
clique tree given as an input.  Based on these algorithms, we can
define a symmetric binary relation between the set of clique trees and
the set of perfect sequences of maximal cliques.  In this article
we consider to describe this relation using a bipartite graph.  We
prove that the bipartite graph is connected for every chordal graph.
This result allows us to
construct a connected Markov chain over the set of clique trees and
the set of perfect sequences of maximal cliques of a given chordal
graph.  The Markov chain is potentially useful for optimizing over the
set of clique trees or over the set of perfect sequences of 
maximal cliques. 
In the proof of the connectedness of the bipartite graph we use 
the induction on the number of maximal
cliques and we can confirm the usefulness of our characterization of
endpoints of clique tree by using the notion of 
boundary cliques. 

Finally we consider the question of uniqueness of clique trees.
As mentioned above a chordal graph may have many clique trees.  As two
extremes, there exists a chordal graph such that an arbitrary tree is a 
clique tree for it and there also exists a chordal graph 
such that the clique tree is unique.  
We derive a necessary and sufficient
condition on chordal graphs for the arbitrariness and for the
uniqueness of their clique trees.

The organization of this paper is as follows. 
In Section 2 we prepare notations and present some preliminary
facts on the simplicial vertices, the clique trees and the perfect
sequences of maximal cliques of a chordal graph. 
In Section 3 we consider boundary cliques and give some characterization 
of endpoints of clique trees in relation to boundary cliques.
We also characterize a final maximal clique in a perfect sequence by
using the notion of boundary cliques.
In Section 4 we define a symmetric binary relation between the set
of clique trees and the set of perfect sequences of maximal cliques and
consider to describe the relation by a bipartite graph. 
In particular we prove that the bipartite graph is connected.
In Section 5 we derive the necessary and sufficient condition for the
arbitrariness and the uniqueness of clique trees.  We end the paper with
some concluding remarks in Section 6.

\section{Preliminaries}
In this section we prepare notations, definitions and some basic
results on chordal graphs required in the subsequent sections.
Throughout this paper we assume that the undirected graph 
$G$ is a {\em connected} chordal graph,
because for a general chordal $G$ it suffices to consider clique trees
separately for each connected component.

\subsection{Notations and definitions}
Let $\vertices$ be the set of vertices in $G$.
Denote  by $\cC$ and $\cS$ the set of  
maximal cliques and the set of minimal vertex
separators in $G$, respectively.
It is well known that $G$ is chordal if and only if every minimal
vertex separator is a clique (Dirac\cite{Dirac}).
Define $n = \vert \vertices \vert$ and $K=\vert \cC \vert$.
For a subset of vertices $\vertices' \subset \vertices$, the subgraph 
induced by $\vertices'$ is denoted by $G(\vertices')$.
Let $\cC(\vertices')=\cC(G(\vertices'))$ 
and $\cS(\vertices')=\cS(G(\vertices'))$ denote the set of the maximal
cliques and the set of minimal vertex separators of $G(\vertices')$. 
For a subset of maximal cliques $\cC' \subset \cC$, denote 
$\vertices(\cC')=\bigcup_{C \in \cC'} C \subseteq V$.  Here a maximal
clique $C$ is
considered to be a subset of $\vertices$.
In this article we use
$\subset$ for a proper containment and $\subseteq$ for a containment with
equality allowed.

For a vertex $\avertex \in \vertices$, 
we denote by $N_G(\avertex)$ the open adjacency set of $\avertex$ in $G$,
i.e.\ the set of all neighbors of $\avertex$ in $G$, and by 
$N_G[\avertex]$ the closed adjacency set of $\avertex$ in $G$,
i.e. $N_G[\avertex]=N_G(\avertex) \cup \{\avertex\}$.
For a subset of vertices $\vertices' \subset \vertices$, define 
$N_G(\vertices')$ and $N_G[\vertices']$ as follows, 
$$
N_G(\vertices') = \bigcup_{\avertex \in \vertices'} N_G(\avertex) \setminus
\vertices', \quad
N_G[\vertices'] = \bigcup_{\avertex \in \vertices'} N_G[\avertex].
$$

A tree $T=(\cC,\cE)$ is called a {\em clique tree} for $G$ if 
for any two maximal cliques $C_1 \in \cC$ and $C_2 \in \cC$ and any
$C_3 \in \cC$ on the unique path in $T$ between $C_1$ and $C_2$ it
holds that  
$$
C_1 \cap C_2 \subseteq C_3.
$$
This is known as the {\em junction property} of $T$.
It is well known that a clique tree exists if and only if $G$ is
chordal (e.g. Buneman\cite{Buneman} and Gavril\cite{Gavril}). 
For two maximal cliques $C_1$ and $C_2$ such that 
$(C_1,C_2) \in \cE$, there exists a minimal vertex separator
$S \in \cS$ such that $C_1 \cap C_2 = S$. Hence each edge of $T$ 
corresponds to a minimal vertex separator(e.g. Ho and Lee\cite{HoLee}).
For a subset $\cC' \subset \cC$, 
denote the subtree of $T$ induced by $\cC'$ by $T(\cC')$. 
If $T(\cC')$ is connected, then
$T(\cC')$ also satisfies  the junction property. 
In this case the induced subgraph $G(\vertices(\cC'))$ is also 
chordal with $\cC(\vertices(\cC'))=\cC'$.  

For a (not necessarily maximal) clique $D$ let
\newcommand{\supercliques}{\uparrow}
\[
\cC_{\supercliques D} = \{ C\in \cC \mid D \subseteq C\}
\]
denote the set of maximal cliques containing  $D$.
Then the junction property can be alternatively expressed that 
$\cC_{\supercliques D}$ induces a connected subtree of $T$ for every
clique $D$.
Let $\tilde {\cal C}$ denote the set of all cliques of $G$.
Kumar and Madhavan\cite{Kumar} showed that it is sufficient to
consider
$\cC_{\supercliques S}$ for each minimal vertex separator $S \in \cS$,
i.e.,
\[
\{\cC_{\supercliques S} \mid S \in \cS\}\qquad \text{and} \qquad
\{\cC_{\supercliques D} \mid D \in \tilde {\cal C}\}
\]
induce the same set of connected subtrees of a clique tree.

As already mentioned, there may be many clique trees for $G$.
Ho and Lee\cite{HoLee} and Kumar and Madhavan\cite{Kumar} provided 
efficient algorithms to enumerate all clique trees. 
Ho and Lee\cite{HoLee} gave the number of the clique trees of chordal
graphs explicitly. 
For $S \in \cS$, let $\Gamma_1,\dots,\Gamma_M$ be the connected
components of $G(\vertices \setminus S)$. 
Define $\cM_S$ and $\cC_{\supercliques S}(\Gamma_m \cup S)$,
$m=1,\dots,M$, by
\begin{equation}
 \label{M_S}
  \cM_S = \{m \mid N_G(\Gamma_m)=S\}, \quad
  \cC_{\supercliques S}(\Gamma_m \cup S)
  = \{C \in \cC \mid C \subseteq \Gamma_m \cup S, C \supset S\}. 
\end{equation}
Let $\cT$ be the set of all clique trees for $G$.
Ho and Lee\cite{HoLee} showed that the number of the clique trees 
for $G$ is expressed by 
\begin{equation}
 \label{number-ct}
\vert \cT \vert =
\prod_{S \in \cS}
\left[
\left(
\sum_{m \in \cM_S}
\vert \cC_{\supercliques S}(\Gamma_m \cup S) \vert
\right)^{\vert \cM_S \vert -2}
\cdot 
\prod_{m \in \cM_S}
\vert \cC_{\supercliques S}(\Gamma_m \cup S) \vert
\right].
\end{equation}
We consider to characterize the chordal graphs from the aspect of the
arbitrariness and the uniqueness of clique trees in Section 5.

Other important characterizations of the clique trees are addressed in 
Bernstein and Goodman\cite{BG} and Shibata\cite{Shibata} etc.

A vertex $\avertex \in \vertices$ is called {\em simplicial} if
$N_G(\avertex)$ is a clique. 
Dirac\cite{Dirac} showed that any chordal graph with at least two
vertices has at least two simplicial vertices and that if the graph is
not complete, these can be chosen to be non-adjacent. 
A bijection $\sigma : \{1,\dots,n\} \rightarrow \vertices$ 
is called a {\em perfect elimination scheme} of vertices of $G$ if 
$\sigma(i)$ is a simplicial vertex in 
$G(\bigcup_{j=i}^n\{\sigma(j)\})$. 
It is well known that $G$ is chordal if and only if $G$ contains
a perfect elimination scheme. The perfect elimination scheme is used to 
determine whether a given graph is chordal.
Linear time algorithms to generate a perfect elimination scheme are
proposed in Tarjan and Yannakakis\cite{Tarjan} and
Golumbic\cite{Golumbic} etc.  

For a maximal clique $C$,  let ${\rm Simp}(C)$ denote the set of
simplicial vertices in $C$ and let ${\rm Sep}(C)$
denote the set of non-simplicial vertices in $C$.
Then $C = {\rm Simp}(C) \cup {\rm Sep}(C)$ is a partition (disjoint
union) of $C$.
As shown below in Lemma  \ref{lemma:H-T},
$$
{\rm Sep}(C) = C \cap \vertices(\cC \setminus \{C\})
    = C \cap \bigcup_{S \in \cS}S.
$$
We call ${\rm Simp}(C)$ the {\em simplicial component} of $C$ and
${\rm Sep}(C)$ the {\em non-simplicial component} of $C$, 
respectively.
We call a maximal clique $C$ {\em simplicial} if
${\rm Simp}(C) \neq \emptyset$.  
Note that for brevity of terminology in this paper 
we simply say ``simplicial clique'' instead of ``simplicial maximal
clique''.

Denote the maximal cliques in $G$ by $C_k$, $k=1,\dots,K$.
Define $\cI = \{1,\dots,K\}$.
For the permutation 
$\pi : \cI \rightarrow \cI$, 
define $H_{\pi(k)}$, $k=1,\dots,K$, 
and $S_{\pi(k)}$, $k=2,\dots,K$, by
\begin{equation}
 \label{def:HRS}
  H_{\pi(k)} = C_{\pi(1)} \cup \cdots \cup C_{\pi(k)}, \quad
  S_{\pi(k)} = H_{\pi(k-1)} \cap C_{\pi(k)},
\end{equation}
respectively.
The sequence of the maximal cliques 
$C_{\pi(1)},C_{\pi(2)},\dots,C_{\pi(K)}$ is a 
{\em perfect sequence of the maximal cliques} if every
$S_{\pi(k)}$ is a clique and there exists $k' < k$ such that 
$S_{\pi(k)} \subset C_{\pi(k')}$ for all $k \ge 2$. 
This is known as the {\em running intersection property} of the sequence.
There exists a perfect sequence of maximal cliques if and
only if $G$ is chordal and 
then $S_{\pi(k)} \in \cS$ for all $k$ and 
\begin{equation}
 \label{eq:separators}
  \{ S_{\pi(2)}, \dots, S_{\pi(K)} \} = \cS,
\end{equation}
where the same minimal vertex separator $S$ may be repeated several
times on the left-hand side (e.g. Lauritzen\cite{Lauritzen}).
Define the {\em multiplicity} $\nu(S)$ of $S\in \cS$ by 
$$
\nu(S) = \#\{k \mid S_{\pi(k)}=S,\;k=2,\dots,K\}.
$$
It is known that $\nu(S)$ does not depend on $\pi$.
It is also known that there exists a perfect sequence such that 
$C_{\pi(1)}=C_k$ for all $k=1,\dots,K$.
We identify the sequence $C_{\pi(1)},C_{\pi(2)},\dots,C_{\pi(K)}$ with 
the permutation $\pi$ for simplicity for the rest of the paper.
Denote the set of perfect sequences of $G$ by $\varPi$.

\subsection{Some basic facts on chordal graphs}

In this subsection we present some basic facts on chordal graphs
required in the following sections in the form of series of lemmas.
Many results of this section are not readily available in the existing
literature.  However they are of preliminary nature and we do not
intend to claim originality of the results of this subsection.  The
readers may skip the proofs of the lemmas and refer
to the lemmas when needed in checking proofs of our main results in
the later sections.

We first state the following fundamental 
property of the simplicial vertices.
\begin{lemma}[Hara and Takemura\cite{H-T-2}]
 \label{lemma:H-T}
 The following three conditions are equivalent, 
 \begin{itemize}
\setlength{\itemsep}{0pt}
  \item[{\rm (i)}] $\avertex \in \vertices$ is simplicial ;
  \item[{\rm (ii)}] there is only one maximal clique $C$ which includes $\avertex$ ; 
  \item[{\rm (iii)}] $\avertex \notin S$ for all $S \in \cS$.
 \end{itemize}
\end{lemma}

Note that from this lemma it follows that
\[
\vertices(\cC \setminus \{C\}) = V \setminus {\rm Simp}(C), \qquad 
\forall C\in \cC.
\]

Next we consider a relation between a beginning part of a perfect
sequence of maximal cliques and a connected induced subtree of a
clique tree. Let $C_{\pi(1)},\dots,C_{\pi(K)}$ be a 
perfect sequence of the maximal cliques. 
For $k < K$, the subsequence $C_{\pi(1)},\dots,C_{\pi(k)}$ also
satisfies the running intersection property.  
Denote $\cC_{\pi(k)}=\bigcup_{i=1}^k \{C_{\pi(i)}\}$.
Then the induced subgraph $G(\vertices(\cC_{\pi(k)}))$ is 
a chordal graph with 
$\cC(\vertices(\cC_{\pi(k)}))=\cC_{\pi(k)}$. 
Therefore we have the following lemma.

\begin{lemma}
 \label{Th:ct_ps}
 Suppose that $\cC' \subset \cC$ and  $\vert \cC' \vert = k$.
 There exists a clique tree such that the induced subtree $T(\cC')$ is
 connected if and only if there exists a perfect sequence $\pi$ such
 that $\cC' = \cC_{\pi(k)}$. 
\end{lemma}
We consider this relation once again in Section 4.


\begin{lemma}
 \label{lemma:connected}  
 If $G$ is not complete,
 then $G(\vertices(\cC \setminus \{C\}))$ is connected.
\end{lemma}

\begin{proof}
 Since $G$ is connected ${\rm Sep}(C)\neq \emptyset$.
 Let $\avertex \in {\rm Sep}(C)$. 
 From (ii) in Lemma \ref{lemma:H-T}, $\avertex$ is contained in at least 
 two maximal cliques. 
 Then $\avertex \in \vertices(\cC \setminus \{C\})$.
 Since the simplicial component is not a separator of
 $G$ from (iii) in Lemma \ref{lemma:H-T}, 
 $G(\vertices(\cC \setminus \{C\}))$ 
 is connected.
\end{proof}

For our proofs it is important to consider ``small'' minimal vertex
separators.  In particular we consider a  minimal vertex separator
$S \in \cS$ which is minimal in 
$\cS$ with respect to the inclusion relation. 
The following lemma concerns minimal vertex separators
which are minimal in $\cS$ with respect to the inclusion relation.
Denote the connected components of $G(\vertices \setminus S)$ by 
$\Gamma_1,\dots,\Gamma_M$.

\begin{lemma}
 \label{lemma:clique}
 Let $S \in \cS$ be minimal in 
 $\cS$ with respect to the inclusion relation.
 Then 
 \begin{itemize}
  \item[{\rm (i)}]
	    $\cC= \bigcup_{m=1}^M \cC(\Gamma_m \cup S)$ is a disjoint
	    union ;
  \item[{\rm (ii)}] if $G(\Gamma_m \cup S)$ is not complete, 
	    then $\cS(\Gamma_m \cup S) \subset \cS$ ; 
  \item[{\rm (iii)}]  there exists a perfect sequence $\pi$ such that 
 the set of the first $\vert \cC(\Gamma_m \cup S) \vert$ maximal cliques  
 is $\cC(\Gamma_m \cup S)$ for every $m=1,\dots,M$. 
 Hence there exists a clique tree such that the subgraph of it induced by 
 $\cC(\Gamma_m \cup S)$ is connected.
 \end{itemize}
\end{lemma}


 

The proof of this lemma is presented in the Appendix. 
With respect to (i) in this lemma, we note that if $S$ is not minimal in
$\cS$ with respect to the inclusion relation, then in general we only
have $\cC \subseteq \bigcup_{m=1}^M \cC(\Gamma_m \cup S)$.

In the remaining two lemmas of this subsection we consider
properties of the set of maximal cliques 
$\cC_{\supercliques S}= \{ C \in \cC \mid C \supset S\} \subseteq \cC$
containing a minimal vertex separator $S$. 

\begin{lemma}
 \label{lemma:app-1}
 Let $S_1, S_2 \in \cS $ be minimal vertex separators.
 If  $S_1 \neq S_2$, then 
 $\cC_{\supercliques S_1} \neq \cC_{\supercliques S_2}$. 
\end{lemma}

\begin{proof}
 Suppose that $\cC_{\supercliques S_1}=\cC_{\supercliques S_2}$.
 Then we have
 $$
 \bigcap_{C \in \cC_{\supercliques S_1}} C  = 
 \bigcap_{C \in \cC_{\supercliques S_2}} C \supseteq S_1 \cup S_2. 
 $$ 
 Since $S_1 \neq S_2$, we can assume $S_2 \setminus S_1 \neq \emptyset$
 without loss of generality.
 Then we have 
 $$
 \bigcap_{C \in \cC_{\supercliques S_1}} C 
 \setminus S_1 =
 \bigcap_{C \in \cC_{\supercliques S_2}} C 
 \setminus S_1 
 \supseteq S_2 \setminus S_1 \neq \emptyset. 
 $$
 Hence there exists $\avertex \in \vertices(\cC_{\supercliques S_1})$ 
  such that 
 $\avertex \in C \setminus S_1$  for all $C \in \cC_{\supercliques S_1}$. 
 This implies that $G(\vertices \setminus S_1)$ is connected.
 This contradicts that $S_1$ is a minimal vertex separator of $G$.
\end{proof}

Define $K_S$ by $K_S=\vert \cC_{\supercliques S} \vert$. 
Then we obtain the following lemma.

\begin{lemma}
 \label{lemma:C_S}\ 
 \begin{itemize}
  \item[{\rm (i)}]  $\cS(\vertices(\cC_{\supercliques S})) 
	      \subseteq \cS$ for every $S \in \cS$. 
  \item[{\rm (ii)}]  If $\vert \cS(\vertices(\cC_{\supercliques S})) \vert = 1$, 
	      then $\nu(S) = K_S -1$. 
  \item[{\rm (iii)}]  If $\vert \cS(\vertices(\cC_{\supercliques S})) \vert \ge 2$, 
	      then $S \subset S'$ for all $S' \neq S$, 
	      $S' \in \cS(\vertices(\cC_{\supercliques S}))$. 
 \end{itemize}
\end{lemma}

\begin{proof}
 (i)
 $\cC_{\supercliques S}$ induces 
 a connected subtree in any clique tree for $G$. 
 Thus there exists a perfect sequence $\pi$ of $\cC$ such that 
 $\{C_{\pi(1)},\dots,C_{\pi(K_S)}\} = \cC_{\supercliques S}$ 
  from Lemma \ref{Th:ct_ps}.
 Then 
 $$
 \cS(\vertices(\cC_{\supercliques S})) 
 = \bigcup_{k=2}^{K_S}\{S_{\pi(k)}\} 
 \subseteq \cS.
 $$ 
 
 (ii)
 Since  $\{C_{\pi(K_S+1)},\dots,C_{\pi(K)}\} = \cC \setminus 
 \cC_{\supercliques S}$, 
 $S_{\pi(k)} \neq S$ for $k > K_S$ from the running intersection
 property. 
 Hence if $\vert \cS(\vertices(\cC_{\supercliques S})) \vert = 1$, 
 then $\nu(S) = K_S -1$. 

 (iii)
 Let $\cC_{\supercliques (S \cup S')}$ be 
 the set of maximal cliques in $\cC_{\supercliques S}$ which
 include $S'$. 
 Then 
 $$
 \bigcap_{C \in \cC_{\supercliques (S \cup S')}} C
 \supseteq S \cup S'. 
 $$
 Hence if $S \setminus S' \neq \emptyset$, 
 $G(\vertices(\cC_{\supercliques (S \cup S')})
 \setminus S')$  
 is connected, which implies that 
 $G(\vertices(\cC_{\supercliques S}) \setminus S')$ is also connected.
 This contradicts that 
 $S' \in \cS(\vertices(\cC_{\supercliques S}))$. 
\end{proof}



\section{Boundary cliques and endpoints in clique trees}
In this Section we first define boundary cliques according to
Shibata\cite{Shibata}.  We also introduce an alternative 
terminology of 
{\it simply separated cliques} and 
{\it simply separated vertices}, which seem to be more descriptive.
Next we characterize endpoints of clique trees 
in their relation to boundary cliques. 


\subsection{Boundary cliques
and their properties}
\label{sec:simply-separated-simplicial}



\begin{definition}
 \label{def:ess}
 A simplicial clique $C$ is a boundary clique
 if there exists a maximal clique $C'$ such that 
 \begin{equation}
  \label{eq:ess}
   \mathrm{Sep}(C) = C \cap C'.
 \end{equation}
 Then $C'$ is called a dominant clique.
 We also call $C$ a simply separated clique,
 ${\rm Simp}(C)$ a simply separated component 
 and  the vertices in 
 ${\rm Simp}(C)$ simply separated vertices.
\end{definition}



\begin{rem}
\label{rem:definition-ss}
If $C$ is not simplicial, then $C={\rm Sep}(C)$ is a maximal clique and 
hence there does not exist a dominant clique for $C$.
Therefore if (\ref{eq:ess}) holds, $\mathrm{Simp}(C)$ has to be non-empty.
It follows that the condition (\ref{eq:ess}) alone guarantees that $C$
is simplicial and that $C$ is a boundary clique.
Because of this fact we simply say ``boundary clique'',
``simply separated clique'' or ``simply separated vertex'' instead of 
``simplicial boundary clique'',
``simply separated simplicial clique'' or ``simply separated
simplicial vertex''.
\end{rem}


We now give two characterizations of boundary cliques. 

\begin{propo}
 \label{lemma:ns-cond} If $G$ is not complete, the
 following three conditions are equivalent, 
 \begin{itemize}
  \item[{\rm (i)}] 
   $C$ is a boundary clique ; 
  \item[{\rm (ii)}] 
		   there exists $S \in \cS$ satisfying
 	     \begin{equation}
 	      \label{eq:ns-cond}
	       \mathrm{Sep}(C) = S;
 	     \end{equation}
  \item[{\rm (iii)}]  $G(\vertices(\cC \setminus \{C\}))$ is a chordal graph
	     with 
	     $\cC(\vertices(\cC \setminus \{C\}))=\cC \setminus \{C\}$. 
 \end{itemize}
\end{propo}

\begin{proof}
 (i) $\Rightarrow$ (ii)\; 
 Suppose that (\ref{eq:ess}) holds. 
 Since $C$ is the only maximal clique which includes $\mathrm{Simp}(C)$, 
 $\mathrm{Sep}(C)$ separates $\mathrm{Simp}(C)$ and 
 $C' \setminus \mathrm{Sep}(C)$. 
 On the other hand, for $D \subset \mathrm{Sep}(C)$, 
 $G((C \cup C') \setminus D)$ is connected. 
 This implies that $\mathrm{Simp}(C)$ and 
 $C' \setminus \mathrm{Sep}(C)$ are connected in 
 $G(V \setminus D)$. 
 Hence $\mathrm{Sep}(C) \in \cS$.

 (i) $\Leftarrow$ (ii)\; 
 Suppose that (\ref{eq:ns-cond}) holds. 
 Then there exist $\avertex \in \mathrm{Simp}(C)$ and 
 $\avertex' \in N_G(S) \setminus C$ 
 such that $S$ is a minimal $\avertex-\avertex'$ separator in $G$. 
 Since $S \cup \{\avertex'\}$ is a clique, there exists a 
 maximal clique $C' \in \cC$ satisfying 
 $C' \supseteq S \cup \{\avertex'\}$.
 Then $C \cap C' = S$.

 (i) $\Rightarrow$ (iii)\; 
 Assume that $C$ satisfies (\ref{eq:ess}). 
 Since 
 $C'' \subseteq \vertices(\cC \setminus \{C\})$
 for all $C'' \in \cC \setminus \{C\}$, 
 we have 
 $\cC \setminus \{C\} 
 \subseteq \cC(\vertices(\cC \setminus \{C\}))
 $.
 Suppose that 
 $\cC \setminus \{C\} 
 \subset \cC(\vertices(\cC \setminus \{C\}))$.
 Then there exists 
 $C'' \in \cC(\vertices(\cC \setminus \{C\}))$ such that 
 $C'' \notin \cC \setminus \{C\}$. 
 If $C'' \nsubseteq C$, $C'' \notin \cC$ is also maximal in $G$.
 This contradicts that $\cC$ is the set of all maximal cliques in $G$.
 Hence $C'' \subset C$. 
 Then 
 $C''$ satisfies $C'' = \mathrm{Sep}(C)$.
 Then from the maximality of $C''$ in $\cC \setminus \{C\}$, there does
 not exist $C' \in \cC \setminus \{C\}$ such that 
 $C' \supseteq \mathrm{Sep}(C)$. 
 This contradicts the assumption that $C$ satisfies 
 (\ref{eq:ess}). 
 Therefore
 $\cC(\vertices(\cC \setminus \{C\}))=\cC \setminus \{C\}$.

 (i) $\Leftarrow$ (iii)\;  
 Suppose that $G(\vertices(\cC \setminus \{C\}))$ is a chordal
 graph with  
 $\cC(\vertices(\cC \setminus \{C\}))=\cC \setminus \{C\}$.
 $G(\vertices(\cC \setminus \{C\}))$ is connected from
 Lemma \ref{lemma:connected} and  
 $C \cap \vertices(\cC \setminus \{C\})$ is a clique. 
 However for all $C' \in \cC \setminus \{C\}$, $C' \nsubseteq C$  
 from the maximality of $C'$.
 Thus 
 $C \cap \vertices(\cC \setminus \{C\})$ 
 is not a maximal clique of $G(\vertices(\cC \setminus \{C\}))$.
 Hence there exists a maximal clique $C' \in \cC \setminus \{C\}$ such that 
 $C' \supset C \cap \vertices(\cC \setminus \{C\})$
 and then $C \cap \vertices(\cC \setminus \{C\})= C \cap C'$. 
\end{proof}

As mentioned in the previous section, any chordal graph which is not 
complete has at least two non-adjacent simplicial components 
(Dirac\cite{Dirac}).   
Shibata\cite{Shibata} showed 
the stronger result that any chordal graph which is not complete has at
least two non-adjacent boundary cliques.
Strengthening this fact, we present the following proposition.
Note that if $G$ is not complete, then $G$ contains at least one 
minimal vertex separator $S$ which is minimal 
in $\cS$ with respect to the inclusion relation.

\begin{propo}
 \label{Th:con-comp}
 Suppose that $S$ is a minimal vertex separator which is minimal 
 in $\cS$ with respect to the inclusion relation.
 Let $\Gamma_1,\dots,\Gamma_M$ be the connected components
 of $G(\vertices \setminus S)$.
 Then each $\Gamma_m$, $m=1,\dots,M$, contains at least one simply
 separated component in $G$.
\end{propo}

\begin{proof}
 It suffices to show that $\Gamma_1$ contains a simply separated
 component.
 $G_1 = G(\Gamma_1 \cup S)$ is also chordal. 
 If $G_1$ is complete, $N_G(\Gamma_1) = S \in \cS$. 
 Hence $\Gamma_1={\rm Simp}(\Gamma_1 \cup S)$ 
 is simply separated also in $G$.

 When $G_1$ is not complete, 
 there exist two non-adjacent simply separated cliques in $G_1$. 
 Since $S$ is a clique, 
 at least one of them does not include $S$. 
 Suppose that $C \in \cC_1=\cC(\Gamma_1 \cup S)$ is simply separated
 in $G_1$ satisfying 
 that the simplicial component of $C$ in $G_1$ does not include $S$.
 Then there exists $C' \in \cC_1$ such that 
 $$
 C \cap \vertices(\cC_1 \setminus \{C\})
 = C \cap C'.
 $$
 From (i) in Lemma \ref{lemma:clique},  $\cC_1 \subset \cC$.
 For all $C_m \in \cC \setminus \cC_1$, $C$ and $C'$ satisfies
 $$
 C \cap C_m \subset C \cap C'.
 $$
 Hence we have
 $$
 C \cap C'
 =
 C \cap \vertices(\cC \setminus \{C\}).
 $$
 Thus $C$ is simply separated also in $G$.
\end{proof}



\begin{figure}[htbp]
 \centering
 \includegraphics{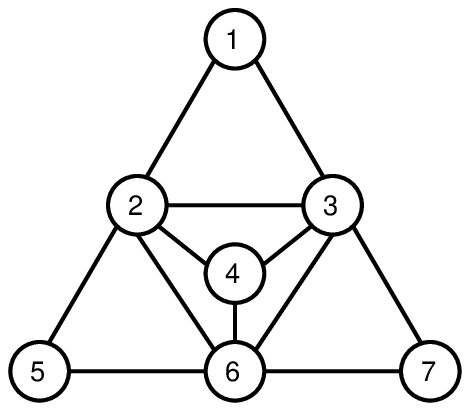}
 \includegraphics{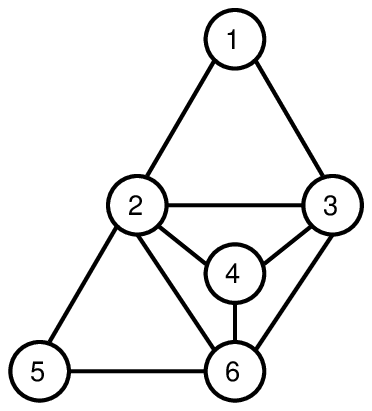}\\
 \hspace*{1cm}(i)\hspace{5.7cm}(ii)
 \caption{Chordal graphs with four and three maximal cliques}
 \label{graph:simplicial-exm}
\end{figure}

\begin{figure}[htbp]
 \centering
 \includegraphics{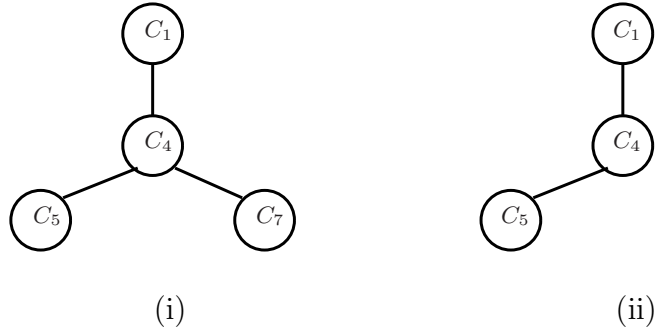}\\
 \hspace*{1cm}(i)\hspace{5.7cm}(ii)
 \caption{The clique trees of the graphs in Figure \ref{graph:simplicial-exm}}
 \label{tree:simplicial-exm}
\end{figure}

Figure \ref{graph:simplicial-exm} presents two graphs with four and three
maximal cliques. 
The set of the simplicial vertices in the graph in
Figure \ref{graph:simplicial-exm}-(i) and (ii) are
$\{1, 4, 5,7\}$ and $\{1, 4, 5\}$, respectively.
Among the simplicial vertices, the vertex 4 is not simply separated in
both \ref{graph:simplicial-exm}-(i) and
\ref{graph:simplicial-exm}-(ii).  
All other vertices are simply separated.
As we will mention in Section 5, 
the clique trees for both of the graphs are uniquely defined as in 
Figure \ref{tree:simplicial-exm}. 
Let $C_\avertex$ 
denote the unique maximal clique containing $\avertex$. 
In the graphs every clique contains a simplicial vertex.
However $C_4$ is not an endpoint in both graphs.


In the literature other classifications of simplicial vertices have
been discussed. 
The class of strongly simplicial vertices are an important subclass of the
simplicial vertices. Following the definition of Agnarsson and
Halld\'{o}rsson\cite{AH}, we define a strongly simplicial component 
as follows. 

\begin{definition}[Strongly simplicial cliques (Agnarsson and
Halld\'{o}rsson\cite{AH})] 
A simplicial clique $C$ is strongly simplicial if
 $$
 \{N_G[\avertex] 
 \mid \avertex \in N_G[{\rm Simp}(C)] \}
 $$
 is a linearly ordered set with respect to the inclusion relation. 
 In this case  
 ${\rm Simp}(C)$ is called a strongly simplicial component 
 and the vertices $\avertex \in {\rm Simp}(C)$ are said to be 
 strongly simplicial.
\end{definition}


If $G$ contains a perfect elimination scheme $\sigma$ such that 
$\sigma(i)$ is a strongly simplicial vertex in 
$G(\bigcup_{j=i}^n\{\sigma(i)\})$, $G$ is called strongly chordal.
The strongly chordal graphs are an important subclass of the chordal
graphs because they yield polynomial time solvability of the domatic set 
and the domatic partition problems. 
Since Farber\cite{Farber} first defined strongly chordal graphs, 
they have been studied by many authors (e.g. Chang and Peng\cite{Chang},
Kumar and Prasad\cite{Kumar-2}).
  

We now show that a strongly simplicial clique is simply separated.


\begin{propo}
 If $C$ is a strongly simplicial clique, then it is simply separated.
\end{propo}

\begin{proof}
 Suppose that
$$
    N_G({\rm Simp}(C)) =
   \{\avertex_1,\avertex_2,\dots,\avertex_m\},
   \quad
   N_G[\avertex_1] \subseteq 
   N_G[\avertex_2] \subseteq \cdots \subseteq 
   N_G[\avertex_m].
$$
 Then 
 \begin{equation}
  \label{adj}
   N_G[\avertex_1] \cap
   N_G[\avertex_2] \cap \cdots \cap
   N_G[\avertex_m] = N_G[\avertex_1].
 \end{equation}
 Since $\avertex_1$ belongs to at least two maximal cliques from (ii) in
 Lemma \ref{lemma:H-T},  
 we have 
 $N_G[\avertex_1] \setminus C \neq \emptyset$.
 Suppose that $\avertex' \in N_G[\avertex_1] \setminus C$.
 From (\ref{adj}) and the fact that 
 ${\rm Simp}(C)$ is simplicial, 
 we have
 $N_G({\rm Simp}(C)) \cap N_G(\avertex') 
= \{\avertex_1,\dots,\avertex_m\}=N_G({\rm Simp}(C))$. 
 Since any vertices in ${\rm Simp}(C)$ and $\avertex'$ are not adjacent
 to each  other,  
 $\{\avertex_1,\dots,\avertex_m\}$ is a minimal $\avertex - \avertex'$
 separator for $\avertex \in {\rm Simp}(C)$.  
\end{proof}

The converse of this proposition does not hold from Figure 1.  Table
\ref{table:simplicial-exm} presents strongly, simply separated and not
simply separated simplicial vertices for the graphs in Figure
\ref{graph:simplicial-exm}.  We see the difference between each
class.

\begin{table}[htbp]
 \centering
 \caption{Simplicial vertices for the graph in Figure 
 \ref{graph:simplicial-exm}}
 \label{table:simplicial-exm}
 \begin{tabular}{lcc}\hline
  & (i) & (ii)\\ \hline
  Strongly simplicial& $\emptyset$ & $1$, $5$\\
  Simply separated but not strongly simplicial & $1$, $5$, $7$ & $\emptyset$\\
  Not simply separated   & $4$ & $4$\\ \hline
 \end{tabular}
\end{table}

\subsection{Relation between the simplicial components and the endpoints of
  clique trees} 

In this section we consider the characterization of endpoints of clique
trees by using the notion of simply separated cliques. 
Shibata\cite{Shibata} showed that if a maximal clique $C$ is an endpoint 
of some clique tree, then it is simply separated.
The following characterization of endpoints of cliques tree includes the
converse of this fact. 

\begin{Th}\ 
\label{If-and-onlyif}
 If $C$ is simply separated, then there exists a
 clique tree $T$ such that $C$ is its endpoint. 
 Furthermore if $C$ and $C'$ are two simply separated cliques in
 two different connected components of 
$G(\vertices \setminus S)$, where $S$ is any minimal vertex separator
which is minimal in $\cS$ with respect to the inclusion relation, then
there exists a  clique tree $T$ such that $C$ and $C'$ are its endpoints.
\end{Th}


\begin{proof}  
 When $K \le 2$, there is nothing to prove.
 Then we assume that $K \ge 3$.

 From Lemma \ref{lemma:connected} and (iii) in Proposition
 \ref{lemma:ns-cond} 
 $G(\vertices(\cC \setminus \{C\}))$ is a connected chordal
 graph with 
 $\cC(\vertices(\cC \setminus \{C\})) = \cC \setminus \{C\}$. 
 Let $T'=(\cC \setminus \{C\}, \cE')$ be a clique tree of 
 $G(\vertices(\cC \setminus \{C\}))$.   
 Denote the dominant clique for $C$ by $C_d$. 
 Consider the tree $T=(\cC,\cE)$, where
 $\cE = \cE' \cup (C,C_d)$.
 Then $C$ is an endpoint of $T$.
 For $C_1 \in \cC \setminus \{C\}$, 
 let $C_2$ be any maximal clique on the unique path between $C$ and $C_1$. 
 From 
 the junction property of $T'$, 
 we have
 $$
 C_2 \supset (C_1 \cap C_d) \supset
 C \cap (C_1 \cap C_d) =
 (C \cap C_1) \cap (C \cap C_d)
 = C \cap C_d.
 $$
 Hence $T$ also satisfies the junction property.

 We move on to prove the second statement.
 Let $S$ a minimal vertex separator which is minimal in
 $\cS$ with respect to the inclusion relation
 and let $\Gamma_1,\dots,\Gamma_M$ be the connected
 components of $G(\vertices \setminus S)$. 
 Let $\Gamma_m \supseteq C$ and $\Gamma_{m'}\supseteq C'$.
 Denote  $S_m = \mathrm{Simp}(C)$ and 
 $S_{m'} = \mathrm{Simp}(C')$, respectively.
 Denote the dominant cliques for $C$ and $C'$ by 
 $C_m$ and $C_{m'}$, respectively. 
 From (ii) in Proposition \ref{lemma:ns-cond}, 
 we have 
 \begin{equation}
  \label{cond:C}
   C \cap C_m = S_m, 
   \quad  
   C' \cap C_{m'} = S_{m'}
 \end{equation}
 Since $C$ and $C'$ belong to different connected components, 
 $C \cap C' \subseteq S$.

 We first consider the case where $C \cap C' \subset S$.
 Then $C \cap C'$ is not a minimal vertex separator of $G$ from
 the minimality of $S$ in $\cS$ with respect to the inclusion relation.
 If $C_m = C'$ then 
 $C' \cap C = C_m \cap C = S_m \in \cS$,
 which is a contradiction. Hence $C_m \neq C'$. 
 Similarly $C' \neq C$.  
 Therefore
 \begin{equation}
  \label{cond:C-C'}
   C_m \in \cC \setminus \{C,C'\}, \quad
   C_{m'} \in \cC \setminus \{C, C'\}.
 \end{equation}
 From (iii) in Proposition \ref{lemma:ns-cond}, 
 $G(\vertices(\cC \setminus \{C\}))$ is a chordal graph with
 $\cC(\vertices(\cC \setminus \{C\})) = \cC \setminus \{C\}$.
 Hence 
 $$
 C' \cap \vertices(\cC \setminus \{C,C'\})
 =C' \cap \vertices(\cC \setminus \{C'\})
 = C' \cap C_{m'}.
 $$
 Thus $C'$ is simply separated also in 
 $G(\vertices(\cC \setminus \{C\}))$. 
 Denote
 $$
 V' =
 \vertices(\cC \setminus \{C,C'\}). 
 $$
 Then $G(V')$ is a chordal graph with 
 $\cC(V') = \cC \setminus \{C,C'\}$ from (iii) in
 Proposition  \ref{lemma:ns-cond}. 
 Hence there exist clique trees for $G(V')$.
 Let $T'=(\cC(V'), \cE')$ be any clique
 tree for $G(V')$.
 Consider the tree $T$ such that 
 \begin{equation}
  \label{tree:simplicial}
   T
   = (\cC, \cE), \quad
   \cE = \cE' \cup \{(C,C_m),(C',C_{m'})\}. 
 \end{equation}
 Then both $C$ and $C'$ are endpoints of $T$ and 
 $T$ can be shown to have the junction property by using the same
 argument as in the proof of the first statement of 
 Theorem \ref{If-and-onlyif}.

 Next we consider the case where $C \cap C' =S$.
 Then from the minimality of $S$ in $\cS$ with respect to the inclusion
 relation, we have 
 $S_m \supseteq S$ and $S_{m'} \supseteq S$. 
 We show the proposition according to the following
 three disjoint cases.
 
 (a) $S_m \supset S$ and $S_{m'} \supset S$. In this case 
 $C_m$ and $C_{m'}$
 in (\ref{cond:C}) satisfy (\ref{cond:C-C'}). 
 Hence there exists a clique tree $T'$ for $G(V')$ and $T$ in 
 (\ref{tree:simplicial}) satisfies the condition of the proposition.

 (b) $S_m \supset S$ and $S_{m'} = S$. In this case $C_m$ in
 (\ref{cond:C}) satisfies $C_m \supset S$, which 
 implies $C' \cap C_m =S$. 
 Hence we can take $C_{m'}$ in (\ref{cond:C}) as $C_{m'}=C_m$.
 $C_{m'} \neq C$ also in this case. 
 Thus there exists a clique tree $T'$ for $G(V')$ in the same 
 way as the above argument.
 Then $T$ in (\ref{tree:simplicial}) satisfies the condition of the
 theorem also in this case. 

 (c) $S_m = S_{m'} = S$. 
 In this case $C$ and $C'$ satisfies
 $$
 C \cap V(\cC \setminus \{C\}) = S, \quad
 C' \cap V(\cC \setminus \{C'\}) = S, 
 $$
 which implies
 $C = \Gamma_m \cap S$ and 
 $C' = \Gamma_{m'} \cap S$.
 From the assumption that $\vert \cC \vert \ge 3$, there exists 
 another connected component $\Gamma_{m''}$ 
 and there exists a maximal clique $C'' \in \cC(\Gamma_{m''} \cup S)$
 such that  
 $$
 C \cap C'' = S, \quad C' \cap C'' = S.
 $$
 Take $C_m$ and $C_{m'}$ in (\ref{cond:C}) as
 $ C_m = C_{m'} = C''$. 
 Then $C_{m'} \neq C$. 
 Hence there exists a clique tree $T'$ for $G(V')$ and $T$ in
 (\ref{tree:simplicial}) satisfies the condition of the theorem 
 also in this case.   
\end{proof}





Combining the results in Shibata\cite{Shibata} and the first statement
of Theorem \ref{If-and-onlyif} we can obtain a necessary and sufficient
condition for a maximal clique to be an endpoint of some clique tree.

\begin{Th}
 There exists a clique tree such that $C \in \cC$ is an endpoint of it
 if and only if $C$ is simply separated. 
\end{Th}



\begin{rem}
 In view of Propositions 
 \ref{Th:con-comp} and  Theorem \ref{If-and-onlyif}
one might ask the 
 following question.  
 Choose  simply separated cliques from each
 connected component: 
 $C_m \in \cC(\Gamma_m \cup S)$, $m=1,\dots,M$.
 Does there exist a clique tree $T$ such that all $C_m$'s are endpoints
 of $T$?   The answer is negative as easily seen from the case
 $|\cS|=1$, since a tree has to contain at least one internal node.
\end{rem}

We present an additional result required in the following section. 
This again concerns minimal vertex separators which are 
minimal in $\cS$ with respect to the inclusion relation. 

\begin{propo}
 \label{Th:endpoint}
 Assume that $G$ is not complete.
 Suppose that $S$ is a minimal vertex separator which is minimal 
 in $\cS$ with respect to the inclusion relation.
 Let $\Gamma_1,\dots,\Gamma_M$ be the connected components
 of $G(\vertices \setminus S)$.
 Denote the set of the endpoints in the clique tree $T$ by $\cL(T)$.
 Then for any $T \in \cT$, there exist at least two $m$ and $m'$ 
 satisfying 
 $$
 \cL(T) \cap \cC(\Gamma_m \cup S) \neq \emptyset, \quad
 \cL(T) \cap \cC(\Gamma_{m'} \cup S) \neq \emptyset.
 $$
\end{propo}

\begin{proof}
 If $\vert \cC(\Gamma_m \cup S) \vert = 1$ for all $m$, 
 then theorem is obvious. 
 Assume that there exists $m$ such that 
 $\vert \cC(\Gamma_m \cup S) \vert \ge 2$.

 Suppose that there exist a clique tree $T \in \cT$ and $m$ satisfying
 $\cL(T) \subseteq \cC(\Gamma_m \cup S)$. 
 Then 
 there exist $C_1 \in \cL(T)$ and 
 $C_2 \in \cL(T)$ 
 such that the path between $C_1$ and $C_2$ contains 
 $C_3 \in \cC$ 
 satisfying 
 $C_3 \in \cC(\Gamma_{m'} \cup S)$, $m' \neq m$.
 This implies that the subtree induced by 
 $\cC(\Gamma_m \cup S)$ of $T$ 
 is disconnected, which contradicts (iii) in Lemma \ref{lemma:clique}.
\end{proof}

\subsection{Some properties of perfect sequences in the context of the
  boundary cliques} 

In the context of boundary cliques, 
perfect sequences are shown to have the following properties.

\begin{Th}
 \label{Th:nscond-ps}
 $C \in \cC$ is simply separated if and only if 
 there exists a perfect sequence $\pi$ such that $C_{\pi(K)}=C$.
\end{Th}

\begin{proof}
 Suppose that there exists a perfect sequence 
 $\pi$ such that
 $C_{\pi(K)}=C$. 
 Then from the running intersection property, 
 there exists $k' \le K-1$ satisfying
 \begin{equation}
  \label{nscond:C_k}
   C \cap (\bigcup_{i=1}^{K-1} C_{\pi(i)}) = C \cap C_{k'}.
 \end{equation}
 Hence $C$ is simply separated.

 Conversely assume that $C$ is simply separated.
 Then there exists a clique tree with an endpoint $C$. 
 Hence there exists a perfect sequence $\pi$ of $\cC \setminus \{C\}$
 from Lemma \ref{Th:ct_ps}.
 Since $C$ is simply separated, there exists $C' \in \cC$ satisfying
 $$
 C \cap V(\cC \setminus \{C\}) = C \cap C'.
 $$
 Thus $C_{\pi(1)},\dots,C_{\pi(K-1)},C_{\pi(K)}$ with 
 $C_{\pi(K)}=C$ is also perfect.
\end{proof}



\begin{lemma}
 \label{lemma:subseq}
 Let $\pi$ be a perfect sequence of $\cC$. 
 Denote $\cC_{\pi(k)}=\{C_{\pi(1)},\dots,C_{\pi(k)}\}$.
 Let $\pi'$ be any perfect sequence of 
 $G(\cC_{\pi(k)})$.
 Then 
 \begin{equation}
  \label{seq}
 C_{\pi'(1)},\dots,C_{\pi'(k)},C_{\pi(k+1)},\dots,C_{\pi(K)}
 \end{equation}
 is a perfect sequence of $\cC$.
\end{lemma}

\begin{proof}
 Since 
 $$
 \{C_{\pi(1)},\dots,C_{\pi(k)}\}=\{C_{\pi'(1)},\dots,C_{\pi'(k)}\},
 $$
 $C_{\pi(k')}$, $k'=k+1,\dots,K$, also satisfy the running intersection
 property in the sequence (\ref{seq}).
\end{proof}

From Theorem \ref{Th:nscond-ps} and Lemma \ref{lemma:subseq}, 
the following corollary is obviously obtained.

\begin{corollary}
 \label{cor:subseq}
 Suppose $C$ is simply separated.
 Let $\pi'$ be any perfect sequence of the maximal cliques for the
 induced subgraph $G(V(\cC \setminus \{C\}))$. 
 Then $C_{\pi'(1)},\dots,C_{\pi'(K-1)},C$ is also perfect.
\end{corollary}

\section{Bipartite graph expression of the relation between the set of
  clique  trees and the set of perfect sequences}
In this section we consider the relation between the set of perfect
sequences and the set of clique trees which we once discussed in Lemma 
\ref{Th:ct_ps}. 
Let $\pi$ be a perfect sequence of $\cC$ and 
$S_{\pi(2)},\dots,S_{\pi(K)}$ be the corresponding minimal vertex
separators defined in (\ref{def:HRS}). 
Lauritzen\cite{Lauritzen} considers the following algorithm to generate 
a tree $T=(\cC,\cE)$ from $\pi \in \varPi$. 
\begin{algorithm}
 \label{ps_to_ct}
 \ \vspace{0.2cm}\\
 \begin{tabular}{lcl}
  Input & : & $\pi \in \varPi$\\
  Output& : & $T = (\cC,\cE)$\\
 \end{tabular}
\vspace{0.2cm}\\
 {\bf begin}\\
 \hspace*{5mm} $\cE \leftarrow \emptyset$\\
 \hspace*{5mm} {\bf for} $k=2$ to $K$ {\bf do}\\
 \hspace*{5mm} {\bf begin}\\
 \hspace*{5mm} \hspace*{5mm} Choose any $k'$ such that $k'< k$ and 
 $S_{\pi(k)} = C_{\pi(k')} \cap C_{\pi(k)}$
 and 
 $\cE \leftarrow \cE \cup \{(C_{\pi(k)},C_{\pi(k')})\}$\\
 \hspace*{5mm} {\bf end}\\
 {\bf end}
\end{algorithm}

As stated in Lauritzen\cite{Lauritzen}, 
any tree generated by Algorithm \ref{ps_to_ct} is a clique tree.
Conversely consider the following algorithm to generate a sequence 
of maximal cliques from a clique tree $T=(\cC,\cE) \in \cT$.

\begin{algorithm}
 \label{ct_to_ps}
 \ \vspace{0.2cm}\\
 \begin{tabular}{lcl}
  Input & : & $T = (\cC,\cE) \in \cT$\\
  Output & : & $\pi$\\
 \end{tabular}
 \ \vspace{0.2cm}\\
 {\bf begin}\\
\hspace*{5mm}
 Choose any $C \in \cC$;\\
 \hspace*{5mm}   Set $C$ as the root and thereby direct all edges in $T$.\\
 \hspace*{5mm} This induces a partial order in $\cC$ such that $C_a \prec C_b$\\
 \hspace*{5mm} if there exists a directed path from $C_a$ to $C_b$.\\
 \hspace*{5mm} Sort $\cC$ topologically according to the order;\\
  \hspace*{5mm} generate $C_{\pi(1)},\dots,C_{\pi(K)}$ with $C_{\pi(1)}=C$;\\ 
 {\bf end}
\end{algorithm}

Lauritzen\cite{Lauritzen} also showed that any sequence generated by
this algorithm is perfect. 
Also by following Algorithm \ref{ps_to_ct} and \ref{ct_to_ps}, 
we can confirm the result of Lemma \ref{Th:ct_ps}.

In the rest of this section we consider the relation between the set
of clique trees and the set of perfect sequences of maximal cliques
through Algorithm \ref{ps_to_ct} and Algorithm \ref{ct_to_ps}.  First
we note the following result.

\begin{lemma}
 \label{Th:symmetric}
 $T \in \cT$ can be generated by by Algorithm \ref{ps_to_ct} with the
 input $\pi \in \varPi$ if and only if $\pi$ can be generated by
 Algorithm \ref{ct_to_ps} with the input $T$. 
\end{lemma}

\begin{proof}
 Suppose that $\pi$ is generated from $T=(\cC,\cE)$ by Algorithm
 \ref{ct_to_ps}. 
 Then for $k \ge 2$, there exists $k' < k$ such that 
 $(C_{\pi(k')},C_{\pi(k)}) \in \cE$.
 $C_{\pi(k)}$ is an endpoint in the subtree 
 $T(\cC_{\pi(k)})$, where $\cC_{\pi(k)}=\{C_{\pi(1)},\dots,C_{\pi(k)}\}$. 
 Hence from the junction property we have
 $$
 C_{\pi(k)} \cap C_{\pi(k')} 
 \supseteq 
 C_{\pi(k)} \cap C_{\pi(k'')}, 
 \quad
 k'' < k, 
 \quad
 k'' \neq k', 
 $$
 which implies
 $$
 C_{\pi(k)} \cap \bigcup_{i=1}^{k-1} C_{\pi(i)} = 
 C_{\pi(k)} \cap C_{\pi(k')}.
 $$
 Hence by Algorithm \ref{ct_to_ps}, we can generate
 $T$ from $\pi$.

 Next we assume that $T$ is generated from $\pi$ by Algorithm
 \ref{ps_to_ct}. 
 Then for any $k \le K$ the induced subtree 
 $T(\cC_{\pi(k)})$ is connected and $C_{\pi(k)}$ is an endpoint of
 $T(\cC_{\pi(k)})$. 
 Set $C_{\pi(1)}$ as the root of $T(\cC_{\pi(k)})$ and 
 consider the directed tree as in Algorithm \ref{ct_to_ps} and 
 denote it by $T(\cC_{\pi(k)},C_{\pi(1)})$. 
 Let $C_{\pi'(1)},\dots,C_{\pi'(k-1)}$ be any sequence which is
 compatible with the order in $T(\cC_{\pi(k)},C_{\pi(1)})$ and satisfies  
 $\pi'(j) \neq \pi(k)$ for all $j \le k-1$.
 Since $C_{\pi(k)}$ is an endpoint of $T(\cC_{\pi(k)})$, 
 the sequence $C_{\pi'(1)},\dots,C_{\pi'(k-1)},C_{\pi(k)}$ is also 
 compatible with the order in $T(\cC_{\pi(k)},C_{\pi(1)})$ for all $k$. 
 This implies that $\pi$ is compatible with the order in 
 $T(\cC,C_{\pi(1)})$.
 Hence $\pi$ can be generated from $\cT$.
\end{proof}

From Lemma \ref{Th:symmetric}, we define the following symmetric binary 
relation  $\cR \subseteq \cT \times \varPi$.

\begin{definition}
\label{def:relation} \quad 
$(T, \pi) \in \cR$ if $T$ can be generated from $\pi$ by 
Algorithm \ref{ps_to_ct} 
\end{definition}

Let $\cT(G_{\cC'})$ be the set of clique trees for 
$G_{\cC'} = G(V(\cC'))$. 
Define $\cT_{\cC'}$ by 
$$
\cT_{\cC'} = \{T \in \cT \mid T(\cC') \mbox{ is connected} \}.
$$
Then we have the following lemma.

\begin{lemma}
 \label{Th:induced-subtree}
 If $\cT_{\cC'} \neq \emptyset$, 
 then $\cT(G_{\cC'})=\{T(\cC') \mid T \in \cT_{\cC'}\}$. 
\end{lemma}

\begin{proof}
 Since $\cC'$ induces a connected component for some clique tree, 
 $G_{\cC'}$ is a chordal graph with 
 $\cC(\vertices(\cC'))=\cC'$. 
 Suppose that $T \in \{T(\cC') \mid T \in \cT_{\cC'}\}$.
 Then we have 
 $T \in \cT(G_{\cC'})$, i.e. 
 $\{T(\cC') \mid T \in \cT_{\cC'}\} \subseteq \cT(G_{\cC'})$. 

 Denote $K' = \vert \cC' \vert$. 
 Let  $\varPi(\cC')$ be the set of perfect sequences of $G_{\cC'}$.
 Let $\cR(\cC') \in \cT(G_{\cC'}) \times \varPi(\cC')$ 
 be the binary relation defined as the above for $G_{\cC'}$. 
 Following Algorithm \ref{ct_to_ps} and Lemma \ref{Th:symmetric}, 
 there exists $\pi' \in \varPi(\cC')$ for any 
 $T' = (\cC', \cE') \in  \cT(G_{\cC'})$ 
 such that 
 $(\pi',T') \in \cR(\cC')$.
 Since $\cT_{\cC'} \neq \emptyset$ from the assumption, 
 there exists $\pi \in \varPi$ such that 
 $\pi(k)=\pi'(k)$ for $k=1,\dots,K'$ from Lemma \ref{Th:ct_ps}. 
 From the running intersection property, 
 there exists $k' < k$ for each $C_{\pi(k)}$ such that 
 $$
 C_{\pi(k)} \cap \bigcup_{i=1}^{k-1}C_{\pi(i)} 
 = C_{\pi(k)} \cap C_{\pi(k')}.
 $$
 Then the tree 
 $$
 T=
 \Bigl(
 \cC,\cE' \cup 
 \bigcup_{k=K'+1}^K
 \bigl\{(C_{\pi(k)},C_{\pi(k')})
 \bigr\} \Bigr)
 $$
 can be generated by Algorithm \ref{ps_to_ct} and then
 $T \in \cT_{\cC'}$, which implies 
 $\cT(G_{\cC'}) \subseteq \{T(\cC') \mid T \in \cT_{\cC'}\}$.  
 Hence we obtain 
 $\cT(G_{\cC'}) = \{T(\cC') \mid T \in \cT_{\cC'}\}$.  
\end{proof}

\begin{figure}[htbp]
 \centering
 \includegraphics{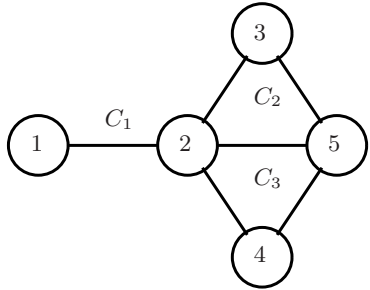}
 \caption{The graph with three cliques}
 \label{graph}
\end{figure}

\begin{figure}[htbp]
 \centering
 \includegraphics{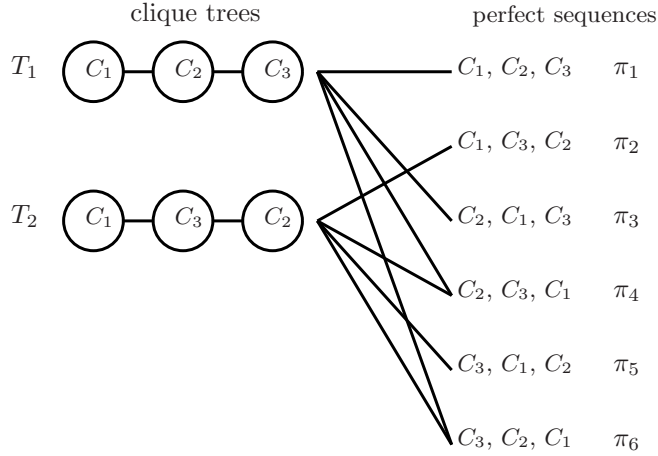}
 \caption{The bipartite graph of the clique trees and the perfect
 sequences for the graph in Figure \ref{graph}}
 \label{exm-bipartite}
\end{figure}

Now we consider to express this binary relations by 
the bipartite graph $B=(\cT \cup \varPi, \cR)$. 
We give a simple example. 
Figure \ref{exm-bipartite} presents the bipartite graph $B$ for 
the graph in Figure \ref{graph}. 
We see that $B$ is not complete. 

In general Algorithm \ref{ps_to_ct} does not necessarily generate every
clique tree if an input perfect sequence is fixed.  
Conversely Algorithm \ref{ct_to_ps} does not necessarily generate every
perfect sequence if an input clique tree is fixed.
Now we denote $\bar {\cal C}_C = \cC \setminus \{C\}$. 
Then the bipartite graph $B$ for the general chordal graph can be shown to
have the following property.  

\begin{lemma}
 \label{lemma:bipartite}
 Suppose that $C \in \cC$ is simply separated.
 Let $\cT_{\bar {\cal C}_C} \subset \cT$ denote the set of clique trees for
 $G$ with an endpoint $C$.  
 Then any two clique trees in $\cT_{\bar {\cal C}_C}$ are connected
 on $B=(\cT \cup \varPi, \cR)$. 
\end{lemma}

\begin{proof}
 We prove it by induction on the number $K=|\cC|$ of the maximal cliques.
 If $K \le 2$, the lemma is obvious. 
 Suppose that $K \geq 3$ and that the lemma holds for all chordal graphs  
 with fewer than $K$ maximal cliques.

 Denote $G_{\bar {\cal C}_C} = G(V(\bar {\cal C}_C))$. 
 First we note that 
 $\cT(G_{\bar {\cal C}_C}) 
 = \{T(\bar {\cal C}_C) \mid T \in \cT_{\bar {\cal C}_C}\}$ 
 from Lemma \ref{Th:induced-subtree}.
 Since $C$ is simply separated,
 there exists a perfect sequence of $\bar {\cal C}_C$ from 
 Theorem \ref{Th:nscond-ps}. 
 Denote the set of such perfect sequences by  
 $\varPi(\bar {\cal C}_C)$. 
 Let 
 $\cR(\bar {\cal C}_C) 
 \in \cT(G_{\bar {\cal C}_C}) 
 \times 
 \varPi(\bar {\cal C}_C)$ 
 be the symmetric binary relation in Definition \ref{def:relation}
 for  
 $G_{\bar {\cal C}_C}$. 
 Let $T$ and $T'$ be any two clique trees in $\cT_{\bar {\cal C}_C}$ 
 and 
 $T(\bar {\cal C}_C)$ and $T'(\bar {\cal C}_C)$ be 
 the subtree of $T$ and $T'$ induced by $\bar {\cal C}_C$.
 From the inductive assumption, 
 $T(\bar {\cal C}_C)$ and $T'(\bar {\cal C}_C)$ are connected on
 the bipartite graph   
 $$
 B(\bar {\cal C}_C) 
 =\bigl(\cT(G_{\bar {\cal C}_C})  \cup 
 \varPi(\bar {\cal C}_C) , 
 \cR(\bar {\cal C}_C)\bigr).
 $$
 Suppose that 
 $$
 T(\bar {\cal C}_C)=\tilde{T}_0,\tilde{\pi}_1,\tilde{T}_1,
 \dots,\tilde{T}_{p-1},\tilde{\pi}_p,\tilde{T}_p=
 T'(\bar {\cal C}_C)
 $$ 
 $$
 \tilde{T}_i = (\bar {\cal C}_C, \tilde{\cE}_i) \in 
 \cT(G_{\bar {\cal C}_C}),  
 \quad i=0,\dots,p,
 $$
 $$
 \tilde{\pi}_i \in \varPi(\bar {\cal C}_C), \quad i=1,\dots,p
 $$
 is a path from 
 $T(\bar {\cal C}_C)$ to $T'(\bar {\cal C}_C)$ on 
 $B(\bar {\cal C}_C)$.
 Since $C$ is simply separated, the sequence \\
 $C_{\tilde{\pi}_i(1)},\dots,C_{\tilde{\pi}_i(K-1)},C$ is also a perfect
 sequence of $G$ for all $i=1,\dots,p$ from Corollary \ref{cor:subseq}
 and denote it by $\pi_i$. 
 Let $C'$ be the maximal clique which is adjacent to $C$ on $T'$.
 Define $T_i$ by 
 $$
 T_0 = T, \quad 
 T_i=(\cC, \tilde{\cE}_i \cup \{(C,C')\}), \quad
 i=1,\dots,p. 
 $$ 
 Since $(\tilde{T}_i,\tilde{\pi}_i) \in \cR(\bar {\cal C}_C)$ and 
 $(\tilde{T}_{i-1},\tilde{\pi}_i) \in \cR(\bar {\cal C}_C)$, 
 we also have $(T_i,\pi_i) \in \cR$ and 
 $(T_{i-1},\pi_i) \in \cR$ from the definition of $T_i$.
\end{proof}



By using these lemmas we can show the connectivity of the bipartite
graph $B$. 

\begin{Th}
 \label{Th:bipartite}
 The bipartite graph $B=(\cT \cup \varPi, \cR)$ 
 for any chordal graph $G$ is connected.
\end{Th}

\begin{proof}
 For any perfect sequence $\pi$ 
 there exists a clique tree $T$ such that $(T,\pi) \in \cR$.
 Hence it suffices to show that any two clique trees are connected
 on $B$. 


 Let $T$ and $T'$ be any two clique trees for $G$. 
 Denote the set of endpoints in $T$ and $T'$ by 
 $\cL(T)$ and $\cL(T')$, respectively.
 Suppose that $S$ is a minimal vertex separator which is minimal in $\cS$
 with respect to the inclusion relation. 
 Let the connected components of $G(\vertices \setminus S)$ be denoted by 
 $\Gamma_1,\dots,\Gamma_M$.
 Then from Proposition \ref{Th:endpoint} there exist maximal cliques 
 $C_a \in \cL(T)$, $C_b \in \cL(T)$, $C_{a'} \in \cL(T')$ and 
 $C_{b'} \in \cL(T')$ such that 
 $$
 C_a \setminus S \subseteq \Gamma_a, \quad
 C_b \setminus S \subseteq \Gamma_b, \quad
 a \neq b,
 $$
 $$
 C_{a'} \setminus S \subseteq \Gamma_{a'}, \quad
 C_{b'} \setminus S \subseteq \Gamma_{b'}, \quad
 a'\neq b'
 $$
 If the maximal cliques satisfy one of the following conditions, 
 \begin{equation}
  \label{cond:clique}
   C_a = C_{a'}, \quad
   C_a = C_{b'}, \quad
   C_b = C_{a'}, \quad
   C_b = C_{b'},
 \end{equation}
 then $T$ and $T'$ are connected on $B$ from Lemma
 \ref{lemma:bipartite}.
 
 Suppose the maximal cliques do not satisfy any of
 the conditions in (\ref{cond:clique}). 
 Since $a'\neq b'$,
 one of $\Gamma_{a'}$ and $\Gamma_{b'}$ is not equal to 
 $\Gamma_a$. We now assume $\Gamma_a \neq \Gamma_{b'}$ without loss of 
 generality. 
 From 
 Theorem \ref{If-and-onlyif} there exists a  clique tree $T''$ 
 such that both $C_a$ and $C_{b'}$ are endpoints of it.
 Then from Lemma \ref{lemma:bipartite}, 
 $T$ and $T''$ are connected and $T''$ and $T'$ are connected.
 Hence $T$ and $T'$ are connected.
\end{proof}
\section{Arbitrariness and uniqueness of the clique trees}
In this section we consider to characterize chordal graphs from the
aspect of the arbitrariness and the uniqueness of its clique trees.
With respect to the arbitrariness of the clique trees, we can obtain the
following result.

\begin{Th}
 \label{Th:arbitrary}
 Let $G$ be a chordal graph with at least two maximal cliques.
 An arbitrary tree with the set of nodes $\cC$ is a clique tree of 
 $G$ if and only if $\vert \cS \vert = 1$.
\end{Th}

\begin{proof}
Suppose that $\vert \cS \vert=1$ and $S \in \cS$.
Hence the only restriction imposed on the clique trees for $G$ is that 
$\cC_{\supercliques S}$ induces a connected subtree.
From (\ref{def:HRS}) and (\ref{eq:separators}), 
we have $\cC_{\supercliques S} = \cC$. 
This implies that an arbitrary tree with the set of nodes $\cC$ 
is a clique tree for $G$. 

Conversely suppose that $\vert \cS \vert \ge 2$.
Let $S_1$ and $S_2$ be any two minimal vertex separators of $G$.
Then $\cC_{\supercliques S_1} \neq \cC_{\supercliques S_2}$ from Lemma \ref{lemma:app-1}.
Hence we can assume 
$\cC_{\supercliques S_2} \setminus \cC_{\supercliques S_1} \neq \emptyset$
without loss of generality and suppose 
$C \in \cC_{\supercliques S_2} \setminus \cC_{\supercliques S_1}$. 
Let $T'$ be a clique tree such that the set of nodes is 
$\cC_{\supercliques S_1} \cup \{C\}$ and $C \notin \cL(T')$. 
Then any clique tree $T$ such that $T'=T(\cC_{\supercliques S_1} \cup \{C\})$
does not satisfy the condition that 
$\cC_{\supercliques S_1}$ induces a connected subtree.
\end{proof}

On the other hand the necessary and sufficient condition for the clique
tree to be unique is given as follows. 

\begin{Th}
 \label{Th:unique}
 The clique tree for $G$ is unique if and only if 
 \begin{itemize}
  \item[{\rm (i)}] $\vert \cS \vert = \vert \cC \vert -1$, i.e. $\nu(S) = 1$
	     for all $S \in \cS$; 
  \item[{\rm (ii)}] Any two minimal vertex separators of $G$ do not have
	     the inclusion relation. 
 \end{itemize}
\end{Th}

In order to prove Theorem \ref{Th:unique}, 
we note the following lemma. 

\begin{lemma}
 \label{ns-cond:C_S=2}
 $\cS$ satisfies the
 conditions {\rm (i)} and {\rm (ii)} in Theorem \ref{Th:unique} if and only if 
 $K_S = \vert \cC_{\supercliques S} \vert = 2$ for all $S$. 
\end{lemma}

\begin{proof}
 Suppose that there exists $S \in \cS$ satisfying $K_S \ge 3$. 
 When $\vert \cS(\vertices(\cC_{\supercliques S})) \vert \ge 2$, 
 $S \subset S'$ for all $S' \neq S$, 
 $S' \in \cS(\vertices(\cC_{\supercliques S}))$ from (iii) in Lemma
 \ref{lemma:C_S}. 
 When $\vert \cS(\vertices(\cC_{\supercliques S})) \vert =1$, 
 $\nu(S) \ge 2$ from (ii) in Lemma \ref{lemma:C_S}.
 Hence $\cS$ does not satisfy (i) or (ii).
 
 Next we suppose that $\vert \cS \vert < \vert \cC \vert -1$.
 Then there exists $S \in \cS$ such that $\nu(S) \ge 2$. 
 From (\ref{def:HRS}), $S$ satisfies 
 $K_S \ge 3$.

 Suppose that there exist $S \in \cS$ and $S' \in \cS$ such that 
 $S \subset S'$. 
 Then it is obvious that $\cC_{\supercliques S} \supseteq
 \cC_{\supercliques S'}$. 
 From Lemma \ref{lemma:app-1}, $\cC_{\supercliques S} \neq
 \cC_{\supercliques S'}$. 
 Hence $\cC_{\supercliques S} \supset \cC_{\supercliques S'}$ and 
 $K_S \ge 3$. 
\end{proof}


 

{\indent\sc Proof of Theorem \ref{Th:unique}. }
Let $T=(\cC, \cE)$ be a clique tree for $G$. 
Suppose that $\cS$ satisfies the conditions (i) and (ii). 
From Lemma \ref{ns-cond:C_S=2}, 
$K_S =2$ for all $S \in \cS$.
Hence the restriction that $\cC_{\supercliques S}$ induces 
a connected subtree 
is equivalent to $\cC_{\supercliques S} \in \cE$, i.e. 
$\{\cC_{\supercliques S} \mid S \in \cS\} \subset \cE$.
The number of restrictions is 
$$
\vert \cS \vert = K-1 = \vert \cE \vert.
$$
Thus $\{\cC_{\supercliques S} \mid S \in \cS\} = \cE$.
Hence $T$ is uniquely defined from the set of restrictions 
$\{\cC_{\supercliques S} \mid S \in \cS\}$.

Next we assume that $T$ is uniquely defined from $G$. 
Then it suffices to show that $\cS$ satisfies (i) and (ii). 
We prove this by induction on the number of maximal cliques. 
When $K=2$, $\cS$ satisfies $\vert \cS \vert =1$. 
Hence $\cS$ obviously satisfies (i) and (ii). 
Assume $K \ge 3$ and $\cS$ satisfies (i) and (ii) for the chordal graphs 
with fewer than $K-1$ maximal cliques.

Let $C$ be an endpoint of $T$. 
From Theorem \ref{cor:subseq}, there exists a perfect sequence 
$\pi \in \varPi$ such that $C_{\pi(K)}=C$. 
Denote $\bar {\cal C}_C = \cC \setminus \{C\}$ and 
$G_{\bar {\cal C}_C}=G(V(\bar {\cal C}_C))$. 
Define $\cT(G_{\bar {\cal C}_C})$, $\varPi(\bar {\cal C}_C)$ and
$\cR(\bar {\cal C}_C)$ in the same way as in the proof of Lemma
\ref{lemma:bipartite}. 
Suppose that the clique trees for
$G_{\bar {\cal C}_C}$ are 
not uniquely defined and let 
$\tilde{T}_1=(\bar {\cal C}_C, \cE_1)$ and 
$\tilde{T}_2=(\bar {\cal C}_C, \cE_2)$ be two clique trees in 
$\cT(G_{\bar {\cal C}_C})$. 
Then there exist 
$\tilde{\pi}_1 \in \varPi(\bar {\cal C}_C)$ 
and $\tilde{\pi}_2 \in \varPi(\bar {\cal C}_C)$ satisfying
$$
(\tilde{T}_1,\tilde{\pi}_1) \in \cR(\bar {\cal C}_C), \quad
(\tilde{T}_2,\tilde{\pi}_2) \in \cR(\bar {\cal C}_C).
$$
From Theorem \ref{If-and-onlyif}, $C$ is simply separated,
Hence both
$$
C_{\tilde{\pi}_1(1)},\dots,C_{\tilde{\pi}_1(k-1)},C \quad
\text{and} \quad 
C_{\tilde{\pi}_2(1)},\dots,C_{\tilde{\pi}_2(k-1)},C
$$
are perfect sequences of $\cC$ from Corollary \ref{cor:subseq} 
and denote them by $\pi_1$ and $\pi_2$, respectively.
From Proposition \ref{lemma:ns-cond} there exist $k_1 < K$, 
$k_2 < K$ and $S \in \cS$ satisfying  
$$
C \cap \bigcup_{k=1}^{K-1}C_{\pi_1(k)} = 
C \cap C_{\pi_1(k_1)} = S,
\quad
C \cap \bigcup_{k=1}^{K-1}C_{\pi_2(k)} = 
C \cap C_{\pi_2(k_2)} = S.
$$
Then
$$
T_1=(\cC, \cE_1 \cup (C,C_{\pi_1(k_1)})), \quad
T_2=(\cC, \cE_2 \cup (C,C_{\pi_2(k_2)})), 
$$
satisfy that $(T_1,\pi_1) \in \cR$ and $(T_2,\pi_2) \in \cR$ 
and that $T_1 \neq T_2$.
Hence both $T_1$ and $T_2$ are clique trees for $G$, which
contradicts the assumption that $T$ is uniquely defined from $G$. 
Thus $\vert \cT(\bar {\cal C}_C) \vert=1$ and denote the unique
tree by   
$\tilde{T}=(\bar {\cal C}_C, \tilde{\cE})$.
Let $\tilde{\pi} \in \varPi(\bar {\cal C}_C)$ be a perfect sequence
satisfying $(\tilde{T}, \tilde{\pi}) \in \cR(\bar {\cal C}_C)$. 
Then $C_{\tilde{\pi}(1)},\dots,C_{\tilde{\pi}(K-1)},C$ is a perfect
sequence of $\cC$ from Corollary \ref{cor:subseq} and denote it by 
$\pi$.

From the inductive assumption 
$\cS(\vertices(\bar {\cal C}_C))$ satisfies
the conditions (i) and (ii). 
Suppose that there exists $S' \in \cS(\vertices(\bar {\cal C}_C))$ 
such that $S' \supseteq S$.
There exist at least two maximal cliques in $\bar {\cal C}_C$ 
which includes $S'$.
Denote two of such maximal cliques by $C_1$ and $C_2$. 
We note that 
$C \cap C_1 = C \cap C_2 = S$.
Then both $T'_1=(\cC,\tilde{\cE} \cup (C_1,C))$ and 
$T'_2=(\cC,\tilde{\cE} \cup (C_2,C))$ satisfy 
$(T'_1,\pi) \in \cR$ and $(T'_2,\pi) \in \cR$,
which contradicts the assumption. Hence there does not exist $S'$ such
that $S' \supseteq S$. 

Suppose that there exists 
$S' \in \cS(\vertices(\bar {\cal C}_C))$
such that $S' \subset S$.
Let $C'$ be an endpoint of $T$ such that $C' \neq C$. 
Then there exist $S'' \in \cS$ such that 
$$
C' \cap V(\bar {\cal C}_{C'}) = S'',
$$
where 
$\bar {\cal C}_{C'} = \cC \setminus \{C'\}$.
If $S'' \subset S \in \cS(\vertices(\bar {\cal C}_{C'}))$, 
there exist at least two clique trees in $G$ by using the same argument 
as the above. 

Consider the case where $S'' \nsubseteq S$. 
Let $\tilde{T}'=(\bar {\cal C}_{C'}, \tilde{\cE}')$ 
be the unique clique tree for $G(V(\bar {\cal C}_{C'}))$. 
Then $\cS(\vertices(\bar {\cal C}_{C'}))$ satisfies the conditions (i)
and (ii). 
We note that $S$, $S'\in \cS(\vertices(\bar {\cal C}_{C'}))$. 
Since $S' \subset S$, $\cC_{\supercliques S'}$ satisfies 
$\cC_{\supercliques S'} \subset \cC_{\supercliques S}$. 
Hence $K_{S'} \ge 3$, which contradicts the fact that 
$\tilde{T}'$ is the unique clique tree for
$G(\vertices(\bar {\cal C}_{C'}))$ and
$\cS(\vertices(\bar {\cal C}_{C'}))$ satisfies the conditions
(i) and (ii). 
Hence there does not exist 
$S' \in \cS(\vertices(\bar {\cal C}_C))$ such that
$S' \subset S$. 
As a result $\cS$ satisfies the conditions (i) and (ii).
\hfill\qed\\

In the context of (\ref{number-ct}), we can obviously obtain the
following result.

\begin{Th}
 \label{Th:unique-2}
 Define $\cM_S$ and $\cC_{\supercliques S}(\Gamma_m \cup S)$ as in
 (\ref{M_S}).  
 Then the conditions {\rm (i)} and {\rm (ii)} in Theorem \ref{Th:unique} is
 equivalent to $\vert \cM_S \vert = 2$ and 
 $\vert \cC_{\supercliques S}(\Gamma_m \cup S) \vert = 1$. 
\end{Th}

With respect to the uniqueness of the clique tree, 
we also obtain the following result.

\begin{Th}
 \label{Th:endpoint-unique}
 Let $T$ be the unique clique tree defined from $G$. Then all 
 maximal cliques which are simply separated
 are the endpoints of $T$.
\end{Th}

\begin{proof}
 Suppose that $C$ is simply separated and that $C$ is not
 an endpoint of $T$. 
 Then there exist at least two maximal cliques which are adjacent to 
 $C$ on $T$. Denote them by $C_1$ and $C_2$. 
 Note that $C \cap C_1 \in \cS$ and $C \cap C_2 \in \cS$.
 Denote $S_1 = C \cap C_1$ and $S_2 = C \cap C_2$. 
 From Proposition \ref{lemma:ns-cond} there exists $S \in \cS$ satisfying
 (\ref{eq:ns-cond}) and hence  
 $S_1$ and $S_2$ satisfy
 $S_1 \subseteq S$ and $S_2 \subseteq S$, respectively. 
 $S_1 = S$ and $S_2 = S$ contradicts (i) in Theorem \ref{Th:unique}
 and $S_1 \subset S$ or $S_2 \subset S$ contradicts (ii) in Theorem
 \ref{Th:unique}. 
\end{proof}

\section{Concluding remarks}
In this article we considered characterizations of the set of clique
trees in three ways.  
In Section 3 we addressed boundary cliques and gave some
characterizations of endpoints of clique trees in relation to 
boundary cliques. 
In Section 4 we defined a symmetric binary relation between the set of
clique trees and the set of perfect sequences of maximal cliques and 
we described the relation using a bipartite graph. 
We showed that the bipartite graph is
connected for any chordal graphs.  
In Section 5 we derived a necessary and sufficient condition for the
arbitrariness and for the uniqueness of their clique trees.

Theorem \ref{Th:bipartite} and Theorem \ref{Th:unique}
are proved by induction on the number of maximal cliques.
In the proof the notions of boundary cliques 
and the symmetric binary relation discussed in Section 4 are essential 
and the usefulness of them were confirmed. 

Boundary cliques may be important from the algorithmic point of view. 
The detection of simply boundary cliques may
contribute to more efficient generation of a  perfect sequence of maximal cliques.
The relation between boundary cliques and the simplicial partition used
in a procedure of the isomorphism detection of chordal graphs in
Nagoya\cite{Nagoya} may be also interesting. 

In Hara and Takemura \cite{H-T-2}, \cite{H-T-3}, we proposed statistical
procedures whose performances depend on the choice of  perfect sequences
of maximal cliques for a given chordal graph.  In this kind of situation
it is desirable to optimize the performance over the set of perfect
sequences.  By the connectedness of the bipartite graph of Section 4
we can construct a connected Markov chain over the set of perfect
sequences and search for the optimum perfect sequence.

By following Theorem \ref{Th:unique}, we see that the non-uniqueness of
clique trees is related to the inclusion relations in $\cS$ and the
multiplicity of minimal vertex separators. 
This fact is important in enumerating all clique trees for a given
chordal graph.
By using this fact, we can provide another algorithm to 
enumerate all clique trees with the lists of maximal cliques and minimal
vertex separators given as inputs.

We have obtained partial results on these problems.
They are left for our future investigations.

\section*{Appendix}
\appendix
\section{Proof of Lemma \ref{lemma:clique}}
\ {\indent\sc Proof of} (i). \quad
 Denote $G_m=G(\Gamma_m \cup S)$.  Suppose that there exists $m$ and 
 $C \in \cC(\Gamma_m \cup S)$ such that $C \cap \Gamma_m = \emptyset$, 
 i.e. $C = S$. 
 From the definition of the perfect sequence 
 $S$ contains at least one minimal vertex separator 
 $S' \in \cS(\Gamma_m \cup S)$ such that  
 $S' \subset S$. 
 Then $S'$ separates $\avertex \in \Gamma_m$ and $S \setminus S'$.
 Since $\Gamma_m \cap \Gamma_{m'} = \emptyset$ for all $m' \neq m$, 
 $S'$ also separates $\avertex$ and any vertices in $\Gamma_{m'}$, 
 which contradicts the minimality of $S$ in $\cS$ with respect to the
 inclusion relation. 
 Hence $C \in \cC(\Gamma_m \cup S)$ satisfies
 $C \cap \Gamma_m \neq \emptyset$ for all $m$. Choose $\avertex_m \in
 C\cap \Gamma_m$.

 Now suppose that there exists $C' \in \cC$ such that $C' \supset C$ for 
 $C \in \cC(\Gamma_m \cup S)$.
 This implies that there exists $m' \neq m$ such that
 $(C' \setminus C) \cap \Gamma_{m'} \neq \emptyset$.  Choose 
 $\avertex_{m'}\in (C' \setminus C) \cap \Gamma_{m'}$.  Both
 $\avertex_m$ and $\avertex_{m'}$ belong to $C'$.
 However this contradicts the fact that 
 $\Gamma_m$ and $\Gamma_{m'}$ are not adjacent to each other 
 for all $m' \neq m$.
 Hence  $\bigcup_{m=1}^M \cC(\Gamma_m \cup S) \subseteq \cC$.
 
 Since $C \setminus S$ is connected for all $C \in \cC$, 
 there exists $m$ such that $C \subset \Gamma_m \cup S$. 
 Noting that $\vertices \supset \Gamma_m \cup S$, 
 if $C$ is a maximal clique in $G$, then 
 $C$ is also a maximal clique in $G_m$. 
 Hence $\bigcup_{m=1}^M \cC(\Gamma_m \cup S) \supseteq \cC$.
 As a result we obtain 
 $\bigcup_{m=1}^M \cC(\Gamma_m \cup S) = \cC$.  

 Also it is easy to see that 
 $\Gamma_m \cap \Gamma_{m'} = \emptyset$ , $m\neq m'$,  implies
 $\cC(\Gamma_m \cup S)\cap \cC(\Gamma_{m'} \cup S)=\emptyset$.

\medskip
{\indent\sc Proof of} (ii). \quad
 Let $S'$ be a minimal vertex separator in $G_m$.
 Then $G((\Gamma_m \cup S) \setminus S')$ is disconnected. 
 This implies that $G(\vertices \setminus S')$ is also disconnected.
 Hence $S'$ is a separator in $G$.
 
 There exist $\avertex \in \Gamma_m \cup S$ and 
 $\avertex' \in \Gamma_m \cup S$ such that $S'$ is the minimal  
 $\avertex - \avertex'$ separator in $G_m$.  
 $\avertex$ and $\avertex'$ are connected in 
 $G((\Gamma_m \cup S) \setminus S'')$ for $S'' \subset S$.
 Since $G((\Gamma_m \cup S) \setminus S'')$ is the induced subgraph of 
 $G(\vertices \setminus S'')$,  $\avertex$ and $\avertex'$ are also connected in 
 $G(\vertices \setminus S'')$. 
 Hence $S'$ is a minimal vertex separator of $G$. 
 We have shown that 
 $\cS(\Gamma_m \cup S) \subseteq \cS$.
 Now since $(\Gamma_m \cup S) \setminus S = \Gamma_m$ is connected, 
 $S \notin \cS(\Gamma_m \cup S)$. 
 Hence $\cS(\Gamma_m \cup S)$ is a proper subset of $\cS$.

\medskip
{\indent\sc Proof of} (iii). \quad
 It suffices to show it for $\Gamma_1$. 
 Denote $K_m = \vert \cC(\Gamma_m \cup S) \vert$. 
 Let $\pi_m$ be a perfect sequence of $\cC(\Gamma_m \cup S)$ such that 
 $C_{\pi_m(1)} \supset S$.
 Let $S_{\pi_m(k)} \in \cS(\Gamma_m \cup S)$, $k = 2,\dots,K_m$ 
 be the corresponding minimal
 vertex separator in $G_m$. 
 Consider the sequence 
 \begin{equation}
  \label{sequence:Gamma}
 C_{\pi_1(1)},\dots,C_{\pi_1(K_1)},
 C_{\pi_2(1)},\dots,C_{\pi_M(K_M)}.
 \end{equation}
 Since $\Gamma_m \cap \Gamma_{m'} = \emptyset$, 
 \begin{equation}
  \label{rip-1}
 C_{\pi_m(1)} \cap
 \left(
 \bigcup_{m' < m}\bigcup_{k' \le K_{m'}} C_{\pi_{m'}(k')}
 \right)
 = 
 (C_{\pi_m(1)} \cap S) \cup 
 \left(
 C_{\pi_m(1)} \cap
 \bigcup_{m' < m} \Gamma_{m'} 
 \right) 
 =
 S \subset C_{\pi_1(1)}
 \end{equation}
 and for $k'\ge 2$
 \begin{align}
  \label{rip-2}
   C_{\pi_m(k)} \cap
   \left(
    \bigcup_{m' < m}\bigcup_{k' \le  K_{m'}} C_{\pi_{m'}(k')} \cup
    \bigcup_{k' < k} C_{\pi_{m}(k')}
	\right)
   &
  =
  (C_{\pi_m(k)} \cap S) \cup
  \left(
  C_{\pi_m(k)} \cap \bigcup_{k' < k} C_{\pi_{m}(k')}
  \right)\\ \notag
  &= 
  C_{\pi_m(k)} \cap 
  \bigcup_{k' < k} C_{\pi_{m}(k')}.
 \end{align}
 Since $\pi_m$ is a perfect sequence of $\cC(\Gamma_m \cup S)$, 
 (i) of this lemma, 
 (\ref{rip-1}) and  (\ref{rip-2}) imply that 
 (\ref{sequence:Gamma}) is a perfect sequence of $\cC$.
 
 Then there exists a clique
 tree such that the subgraph of it induced by $\cC(\Gamma_m \cup S)$ is
 connected from Lemma \ref{Th:ct_ps}. 
\hfill\qed
\section{List of notation}
\begin{tabular}{lcll} 
\hline
 $G$ &:& a connected chordal graph\\
 $V$ &:& the set of vertices in $G$\\
 $G(V')$ &:& the subgraph of $G$ induced by $V' \subset
 V$\\
 $\cC$ &:& the set of maximal cliques in $G$\\
 $\cC(V')$ &:& the set of maximal cliques in $G(V')$\\
 $K$ &:& the number of maximal cliques $\vert \cC \vert$\\
 $V(\cC')$ &:& $\displaystyle{\bigcup_{C \in \cC'}C}$ for 
 $\cC' \subset \cC$\\ 
 $\cS$ &:& the set of minimal vertex separators in $G$\\
 $\cS(V')$ &:& the set of minimal vertex separators in $G(V')$\\
 $\nu(S)$ &:& the multiplicity of $S \in \cS$\\
 $\cT$ &:& the set of clique trees for $G$\\
 $\cT(G(V'))$ &:& the set of clique trees for $G(V')$ \\
 $T(\cC')$ &:& the subtree of $T \in \cT$ induced by $\cC' \subset \cC$\\
 $\varPi$ &:& the set of perfect sequences of $\cC$ \\
 $\varPi(\cC')$ &:& the set of perfect sequences of $\cC'$ \\
 \hline
 $G_{\cC'}$ &:& $G(V(\cC'))$ & Sec 4, 5\\
 $V_{m,m'}$ &:& $V(\cC \setminus \{C_m,C_{m'}\})$, 
 $C_m$, $C_{m'} \in \cC$ & Prop. \ref{Th:endpoint}\\
 $\cC_{\supercliques S}$ &:& $\{C \in \cC \mid C \supset S, S \in \cS\}$
 & Sec 2, 5\\
 $\bar {\cC}_C$ &:& $\cC \setminus \{C\}$ & Sec 4, 5\\
 $N_G(v)$  &:& the open adjacency set of $v \in V$ in $G$ & Sec 2, 3\\
 $N_G[v]$  &:& the closed adjacency set of $v \in V$ in $G$ & Sec 3\\
 $N_G(V')$ &:& 
 $\displaystyle{\bigcup_{v \in V'}
 N_G(v) \setminus V'}$ & Sec 3\\
 $N_G[V']$ &:& $\displaystyle{\bigcup_{v \in V'}
 N_G[v]}$ & Sec 3\\
 $\Gamma_1,\dots,\Gamma_M$ &:& the connected components of 
 $G(V \setminus S)$ & Sec 2, 3, 4\\ 
 $G_m$ &:& $G(\Gamma_m \cup S)$ & Appendix A\\
 $\cM_S$ &:& $\{m \mid N_G(\Gamma_m)=S\}$ & (\ref{M_S}),
 (\ref{number-ct}), Th. \ref{Th:unique-2}\\
 $\cC_{\supercliques S}(\Gamma_m \cup S)$ &:& 
 $\{C \in \cC \mid C \subseteq \Gamma_m \cup S, C \supset S\}$
 & (\ref{M_S}), (\ref{number-ct}), Th. \ref{Th:unique-2}\\
 $K_S$ &:&  $\vert \cC_{\supercliques S} \vert$ & Lemma \ref{lemma:C_S}\\
 $K_m$ &:&  $\vert \cC(\Gamma_m \cup S) \vert$ & Appendix A\\
 $\mathrm{Simp}(C)$ &:& the simplicial component in $C \in \cC$ & Sec 2,
 3\\
 $\mathrm{Sep}(C)$ &:& the non-simplicial component in $C \in \cC$ & Sec
 2, 3\\
 $\cT_{\cC'}$ &:& $\{T \in \cT \mid T(\cC') \text{ is connected}\}$ & 
 Lemma 4.2, 4.3\\
 $\cL(T)$ &:& the set of endpoints in $T \in \cT$ & Prop. \ref{Th:endpoint}, 
 Th. \ref{Th:bipartite}\\
 $\cC_{\pi(k)}$ &:& $\{C_{\pi(1)},\dots,C_{\pi(k)}\}$ for $\pi \in \varPi$&
 Th. \ref{Th:ct_ps}, Lemma \ref{lemma:subseq}\\
 $\cR$ &:& the symmetric binary relation on $\cT \times \varPi$
 & Def. 4.1, Sec. 4, 5\\
 $\cR(\cC')$ &:& the symmetric binary relation on
 $\cT(G_{\cC'}) \times \varPi(\cC')$ & Sec. 4, 5\\
\hline
\end{tabular}

\end{document}